\documentclass{article}

\PassOptionsToPackage{numbers, sort&compress}{natbib}

\usepackage[final]{neurips_2024}

\usepackage[utf8]{inputenc} 
\usepackage[T1]{fontenc}    
\usepackage{hyperref}       
\usepackage{url}            
\usepackage{booktabs}       
\usepackage{amsfonts}       
\usepackage{nicefrac}       
\usepackage{mathtools}
\usepackage{microtype}      
\usepackage{multirow}
\usepackage{physics}
\usepackage{xcolor}         
\usepackage{amsmath}
\usepackage{amssymb}
\usepackage{graphicx}
\usepackage{enumerate}
\usepackage{enumitem}
\usepackage{bm}
\newcommand{\bftheta}{{\boldsymbol{\theta}}}

\newcommand{\bfx}{{\boldsymbol{x}}}
\newcommand{\bfz}{{\boldsymbol{z}}}

\newcommand{\bfk}{{\boldsymbol{k}}}

\newcommand{\boldmu}{\boldsymbol{\mu}}
\newcommand{\boldalpha}{\boldsymbol{\alpha}}

\definecolor{hypercolor}{RGB}{174, 60, 60} 
\hypersetup{
  linkcolor  = hypercolor,
  citecolor  = hypercolor,
  urlcolor   = hypercolor,
  colorlinks = true
}

\title{Mean-Field Simulation-Based Inference for \\ Cosmological Initial Conditions}

\author{
    Oleg Savchenko \\
  GRAPPA Institute \\
  University of Amsterdam, The Netherlands \\
  \texttt{o.savchenko@uva.nl} \\
  \And
    Florian List  \\
  Department of Astrophysics \\
  University of Vienna, Austria\\
  \texttt{florian.list@univie.ac.at} \\
  \And
  Guillermo Franco Abell\'{a}n \\
  GRAPPA Institute \\
  University of Amsterdam, The Netherlands \\
  \texttt{g.francoabellan@uva.nl} \\
  \And
  Noemi Anau Montel \\
  GRAPPA Institute \\
  University of Amsterdam, The Netherlands \\
  \texttt{n.anaumontel@uva.nl} \\
  \And
  Christoph Weniger \\
  GRAPPA Institute \\
  University of Amsterdam, The Netherlands \\
  \texttt{c.weniger@uva.nl} \\
}

\begin{document}
\maketitle

\begin{abstract}  
  Reconstructing cosmological initial conditions (ICs) from late-time observations is a difficult task, which relies on the use of computationally expensive simulators alongside sophisticated statistical methods to navigate multi-million dimensional parameter spaces. We present a simple method for Bayesian field reconstruction based on modeling the posterior distribution of the initial matter density field to be diagonal Gaussian in Fourier space, with its covariance and the mean estimator being the trainable parts of the algorithm. Training and sampling are extremely fast ({\it training:} $\sim 1 \, \mathrm{h}$ on a GPU, {\it sampling:} $\lesssim 3 \, \mathrm{s}$ for 1000 samples at resolution $128^3$), and our method supports industry-standard (non-differentiable) $N$-body simulators. We verify the fidelity of the obtained IC samples in terms of summary statistics.
\end{abstract}

\section{Introduction}
\label{intro}
The cosmic initial conditions -- minuscule matter density perturbations seeded 14 billion years ago -- gradually evolved under gravity into the intricately woven filamentary structures of galaxies observed in the Universe around us (e.g.\ \cite{2014Natur.513...71T, 2014MNRAS.438.3465T}). Modeling this process of cosmic structure formation by means of $N$-body simulations is, in principle, relatively straightforward (albeit computationally demanding) -- at least when neglecting all non-gravitational effects
(see \cite{2022LRCA....8....1A, 2020NatRP...2...42V} for recent reviews).

An important task in cosmology addresses the inverse problem: {\it given (simulated or observed) cosmological present-day observations, can the cosmic ICs be reconstructed}? Given a late-time matter density field without velocity information, this problem is ill-posed due to the non-linear nature of gravity, and the IC reconstruction is not unique \cite{Nusser1992, Frisch2002, Crocce_2006}. Probabilistic (Bayesian) reconstruction techniques have successfully been employed \cite{Jasche2013, Jasche_2019}; however, they rely on Hamiltonian Monte Carlo, which is slow due to the large number of parameters (= voxels) $10^6 \lesssim d \lesssim 10^7$ and the complexity of the forward model (which needs to be differentiable such that its gradient can be computed).

Recently, simulation-based inference (SBI, e.g.\ \cite{Cranmer_2020}) has gained significant traction in the astrophysics and cosmology communities, driven by the recent advancements in machine learning  (e.g. \cite{Ho:2024whi, Mishra-Sharma:2021oxe, Montel:2022fhv, 2021arXiv210910360V, Brehmer:2019jyt, Zhao:2021ddh, Cole:2021gwr,Coogan:2022cky,Alvey:2023pkx,SimBIG:2023ywd,DES:2023qwe,Tucci:2023bag,2024arXiv240314750A,2024arXiv240415402V, jeffrey2020solving, Makinen:2021nly, Modi:2023llw}. SBI allows tackling inverse problems in situations where a forward model is available which describes the mapping from input parameters (here: cosmic ICs) to observations (here: late-time matter densities), whereas explicitly formulating a probabilistic model may be hard or even impossible. Equipped with $N$-body simulations as an accurate forward model, SBI ideally lends itself for the task of probabilistic cosmic IC reconstruction.

\paragraph{Our contribution.}
We present a fast (both in terms of training and sampling), simple, yet effective SBI-based technique for drawing cosmic IC posterior samples constrained on late-time density fields. The key components are 1) a trainable maximum a-posteriori (MAP) estimator $\mathbf{\hat{\boldmu}}_\bftheta$ and 2) a trainable likelihood covariance matrix $(\bm Q_\bftheta^L)^{-1}$. Specifically, we model the posterior as Gaussian with a diagonal covariance matrix in Fourier space, which enables near-instantaneous sampling.

\paragraph{Related work.}
Methods for inferring cosmological ICs from late-time observations can generally be subdivided into deterministic and probabilistic frameworks. The former class reconstructs a single `best-fit' IC field and relies either on traditional (analytic and numerical) techniques
\cite{weinberg1992, Nusser1992, Gramann1993, croft1997reconstruction, Frisch2002, Brenier2003, 2017PhRvD..96b3505S, Feng_2018}
or, more recently, deep learning approaches (mostly involving U-Nets) \cite{2023arXiv230313056J, Shallue2023, 2024JCAP...02..031F, 2024SCPMA..6719513W, Doeser:2023yzv}, recurrent inference machines \cite{modi2021cosmicrim}, and variational self-boosted sampling \cite{Modi2023}. While probabilistic reconstruction techniques have traditionally employed Bayesian modeling paired with Hamiltonian Monte Carlo sampling \cite{Jasche2013, Jasche_2019}, score-based generative models have recently emerged as an accurate and much faster contender \cite{legin2023posterior, Park:2023ync, Ono:2024jhn}, and see \cite{List:2023jwo} for a first exploratory study based on SBI alongside autoregressive modeling.

\section{Methodology}
\label{methods}
\paragraph{Simulation data.}
The simplicity of our method presented below does not demand the simulator to be differentiable, which, in combination with the fact that a relatively small training set is sufficient, does not require us to rely on approximation schemes in our forward model and enables the use of full industry-standard cosmological $N$-body codes. As our training data, we employ 2000 pairs of initial (at redshift $z = 127$) and present-day ($z = 0$) matter overdensity fields from the \texttt{Quijote} $N$-body simulations suite \cite{Villaescusa-Navarro:2019bje} with varying random phases of ICs and fixed fiducial Planck cosmology \cite{Planck:2018vyg}. These simulations evolve $512^3$ collisionless particles with the \texttt{Gadget-III} TreePM code \cite{Springel:2005mi} in a periodic cubic volume of $(1 \ \mathrm{Gpc} / h)^3$. The overdensity fields are computed by interpolating the particles onto a grid with resolution of $d = 128^3$, in total amounting to a $\sim 10^6$-dimensional parameter space for the inference.

\paragraph{Statistical model.}

The mentioned ill-posedness of the comological ICs reconstruction makes it natural to formulate the problem in the probabilistic Bayesian setting:
\begin{equation}
\label{bayes}
    p(\bfz|\bfx)=\frac{p(\bfx|\bfz)}{p(\bfx)} p(\bfz),
\end{equation}
where $\bm x\in \mathbb{R}^{d}$ is the observed final overdensity field, and $\bm z\in \mathbb{R}^{d}$ are its ICs, whose distribution for a given observation $\bfx_\mathrm{obs}$ we aim to infer. From observations of the Cosmic Microwave Background \cite{Planck:2018vyg}, we know that the early-universe matter overdensity fluctuations had small amplitudes and to a very high precision can be described as a Gaussian random field; therefore, we can write down the prior as $p(\bfz) \propto \exp{-\frac12 \bfz^T \bm Q^P \bfz}$, where the precision matrix $\bm Q^P$ is diagonal in Fourier space: $\bm Q^P = \mathcal{F}^{\dagger} \bm D^P \mathcal{F}$, with $\bm D^P = 1/P(k)$ determined by the linear matter power spectrum $P(k)$ at $z = 127$, $k:=|\bm k|$ being the wave vector $\bm k$ modulus, and $\mathcal{F}$ denoting the Fourier transform\footnote{In practice, we make use of the closely related discrete Hartley transform rather than the discrete Fourier transform because it has the advantage that it maps real values to real values, while still being unitary and diagonalizing the precision matrix {\cite{Bracewell:83}}.}.

A crucial component of any Bayesian cosmological reconstruction framework, which determines its descriptive power, is the complexity level of the likelihood model. In SBI, the likelihood is implicitly encoded in the forward model, for which reason it is typically not explicitly prescribed, but rather accessed from the forward model by a neural network. In this work, however, we adopt a relatively simple approach, in which we explicitly expand the log likelihood up to quadratic terms w.r.t.\ the ICs $\bfz$, with a diagonal likelihood precision matrix in Fourier space. We found this to be sufficient to yield tight and statistically consistent posterior distributions within the selected scale range, while vastly reducing the complexity and computational cost of our method. On sufficiently large scales, this level of modeling extracts the full information content in the data, as all Fourier modes evolve independently in the linear regime. On smaller scales, our approach neglects information from the cross-correlations between Fourier modes (however, this information is leveraged when estimating the MAP, see below), which should typically lead to conservative, rather than overconfident, posterior estimates. Thus, the likelihood we use is defined as follows:

\begin{equation}
\label{likelihood}
    p(\bfx|\bfz) \propto \exp{-\frac12 (\bfz - \mathbf{\hat{\bfz}}_\bftheta(\bfx))^T \bm Q_\bftheta^L (\bfz- \mathbf{\hat\bfz}_\bftheta(\bfx))},
\end{equation}
where $\bm Q_\bftheta^L = \mathcal{F}^{\dagger} \bm D_{\bftheta}^L \mathcal{F}$ with $\bm D_{\bftheta}^L$ being diagonal, $\mathbf{\hat{\bfz}}_\bftheta(\bfx)$ is the $\bfx$-dependent maximum likelihood estimator (MLE), and the subscript $\bftheta$ denotes quantities which depend on the set of trainable parameters $\bftheta$.

Making use of Bayes' theorem (\ref{bayes}) with the likelihood defined in (\ref{likelihood}), after rearranging the terms one obtains the following form of the normalized posterior distribution:
\begin{equation}
\label{posterior}
    p(\bfz|\bfx) = \cfrac{\exp{-\frac12 (\bfz - \mathbf{\hat{\boldmu}}_\bftheta(\bfx))^T (\bm Q^P + \bm Q_\bftheta^L )(\bfz - \mathbf{\hat{\boldmu}}_\bftheta(\bfx))}}{\sqrt{(2\pi)^d \det(\bm Q^P + \bm Q_\bftheta^L)^{-1}}} \, ,
\end{equation}
where $\mathbf{\hat{\boldmu}}_\bftheta(\bfx)$ is the MAP estimator related to the MLE by $\mathbf{\hat{\boldmu}}_\bftheta(\bfx) = (\bm Q^P + \bm Q_\bftheta^L)^{-1} \bm Q_\bftheta^L \, \mathbf{\hat\bfz}_\bftheta(\bfx)$. 
Motivated by the linearity of cosmic structure growth on large scales, we model the MAP estimator $\mathbf{\hat{\boldmu}}_\bftheta(\bfx)$ as a combination of two terms, where the large-scale regime is captured by a simple multiplicative scaling factor $\boldalpha_\bftheta(\bfk)$, while the complex small-scale evolution is described by a 3D U-Net type \cite{ronneberger2015u} neural network in the following way: 
\begin{equation}
\label{map}
    \mathbf{\hat{\boldmu}}_\bftheta(\bfx) = \mathcal{F}^{\dagger} \left\{ \boldalpha_\bftheta(\bfk) \odot \left[\mathcal{F}\{\bfx\} + \sigma_{> k_\Lambda}(\mathcal{F}\{\operatorname{U-Net}_\bftheta(\bfx)\})\right] \right\}.
\end{equation}
Here, $\sigma_{> k_\Lambda}$ 
denotes a sigmoidal high-pass filter centered at the cut-off scale $k_\Lambda$, which we conservatively chose as $k_{\Lambda} = 0.03 \, h/\mathrm{Mpc}$, $\boldalpha_\bftheta(\bfk)$ is a trainable multiplier for each wave vector $\bfk$, and $\odot$ is the Hadamard (elementwise) product. We found that including $\boldalpha_\bftheta(\bfk)$ slightly improves the results while maintaining essentially the same speed of training. We implemented the U-Net from the \texttt{map2map} package \citep{Jamieson:2022lqc}, slightly simplifying it by reducing the number of hidden channels from 64 to 16. To facilitate the training, before passing the $z=127$ fields to the network, we found it beneficial to divide them by the linear growth factor $D(z=127)$, effectively making the network learn the residuals from the linear theory prediction.

Our training objective is to minimize the negative log posterior probability under the ansatz (\ref{posterior}):
\begin{equation}
    \mathcal{L} = \frac12 \sum_{i=1}^N \left\{(\bfz_i - \mathbf{\hat{\boldmu}}_\bftheta(\bfx_i))^T \bm Q_\bftheta \,(\bfz_i - \mathbf{\hat{\boldmu}}_\bftheta(\bfx_i))\right\}
    - \frac{N}{2} \tr \log \bm Q_\bftheta, 
\end{equation}
where $N$ is the number of samples in the training set, $\bm Q_\bftheta = \bm Q^P + \bm Q_\bftheta^L$, and we used the relation $\log \det \bm Q_\bftheta = \tr \log \bm Q_\bftheta$, 
also leaving out an unimportant constant additive term. Altogether, the trainable set $\bftheta$ of our model consists of the MAP estimator $\mathbf{\hat{\boldmu}}_\bftheta$ parameters (see Eq.\ \ref{map}) and the diagonal part of the likelihood precision matrix $\bm D_{\bftheta}^L$ (in order to ensure the positivity of the latter, we store the respective square root values as the parameters, and square them at inference). Due to its parallels with approximate models encountered in the contexts of statistical physics (e.g.\ \cite[Chap.\ 4]{yeomans1992statistical}) and variational inference \cite{Blei_2017}, we refer to our method as {\it `mean-field simulation-based inference'}. 
While our method, which focuses on the first two moments (mean and variance), is conceptually aligned with the SBI framework outlined in \cite{jeffrey2020solving}, which employs a hierarchy of networks to estimate successive moments of the posterior, our approach differs in both the training objective and the learning strategy. Moreover, our approach is closely connected to uncertainty quantification methods that optimize the Gaussian negative log-likelihood loss, as discussed in \cite{Nix_1994}.

\paragraph{Training procedure.}

We train the model using a batch size of 8 on a single 40 GB \texttt{Nvidia A100} GPU, starting with an initial learning rate of $10^{-2}$ and decreasing it by a factor of 10 after each epoch where the validation loss does not decrease. Training continues until no further improvement is observed in the validation loss, which, under these conditions, takes approximately 30 epochs and 1.5 hours %15 epochs and 1 hour
to converge. The dataset was randomly split with 80\% assigned to training and the remaining 20\% reserved for validation. As a wrapper for our training implementation, we use the \texttt{PyTorch Lightning}-based \cite{PyTorch_Lightning_2019} code \texttt{swyft} \cite{Miller_2021, Miller:2022shs}.

\section{Results}
\label{results}

\begin{figure}[]
    \centering
    \begin{minipage}{0.65\textwidth}
        \centering
        \includegraphics[width=\linewidth]{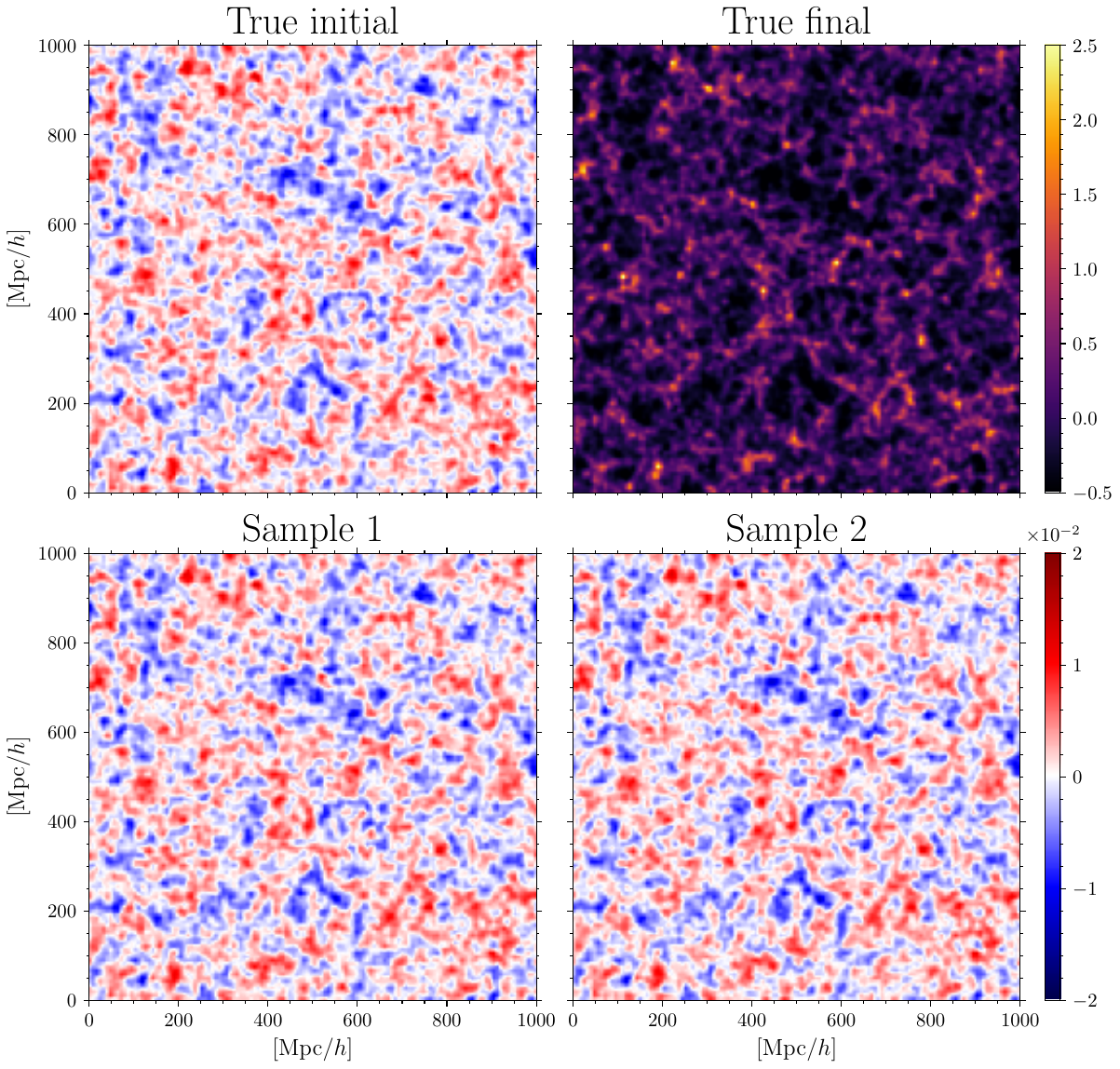}
    \end{minipage}
    \hfill
    \begin{minipage}{0.34\textwidth}
        \centering
        \includegraphics[width=\linewidth]{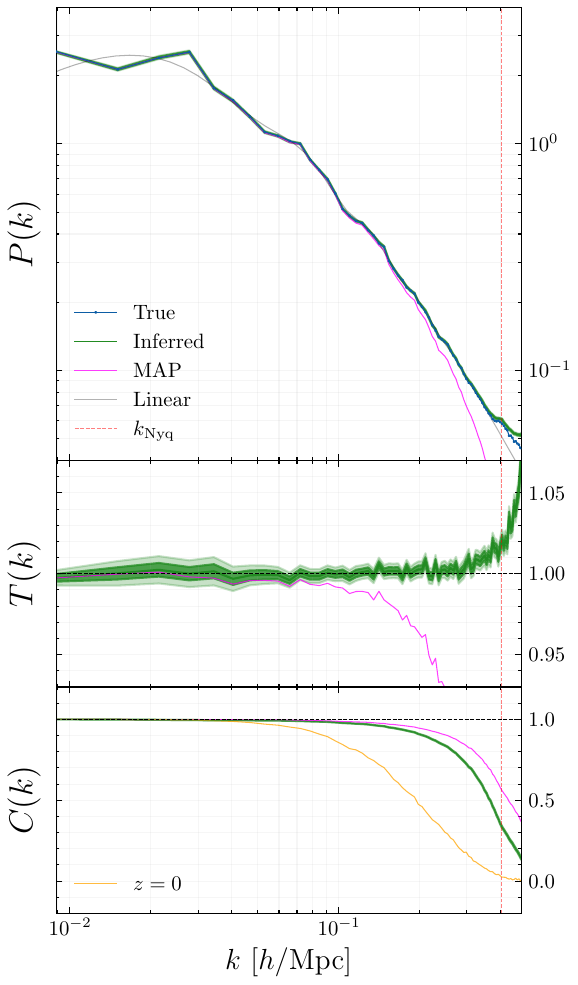}
    \end{minipage}
    \caption{Our method produces posterior IC samples constrained by a given late-time density field ({\it training:} $\simeq 1.5 \, \mathrm{h}$ on a GPU, {\it sampling:} $< 3 \, \mathrm{s}$ for 1000 samples at resolution $128^3$). \textit{Top row}: slices of the initial and final overdensity fields of the target simulation. \textit{Bottom row}: two examples of the generated IC samples. All the shown slices are averaged over the depth of $100 \ \text{Mpc}/h$ in the third axis direction. \textit{Rightmost column}: power spectrum, transfer function, and cross-correlation of the generated samples and the ground truth. Shaded regions correspond to $1\sigma$ and $2\sigma$ errors, and the yellow $C(k)$ line corresponds to the cross-correlation between the final and the initial density fields. Note that the MAP estimate, which is the only quantity recovered by most other approaches, loses significant power at small scales and is not fully able to describe the field's statistical properties.
    \label{fig:results}
    }
\end{figure}

The fact that the prior and the likelihood (and hence also the posterior) are Gaussian and diagonal in the same Fourier basis allows us to obtain samples from the posterior in a very fast and simple way after the training is done. We demonstrate our reconstruction method for one of the fiducial \texttt{Quijote} simulations $\left\{\bfz_{\mathrm{truth}}, \bfx_{\mathrm{obs}}\right\}$ not contained in the training set (see Fig.\ \ref{fig:results}). Given such an observation of the final density field $\bfx_{\mathrm{obs}}$, one evaluation of the MAP estimator (\ref{map}) is required to obtain the mean of the field $\mathbf{\hat{\boldmu}}_\bftheta(\bfx_{\mathrm{obs}})$ ($\lesssim 1 \, \mathrm{s}$ computation time), and having the $\bm Q_\bftheta = \bm Q^P + \bm Q_\bftheta^L$ matrix then allows us to obtain over a thousand %of 
posterior samples in a batched fashion within three seconds on a single GPU, which is orders of magnitude faster than existing methods (see App.~\ref{appendix:map_and_std} for details on MAP estimation and the uncertainty of the modes reconstruction). To confirm that the generated samples have the correct statistical properties, we compute the power spectrum {$P(k)$}, transfer function {$T(k)$} and cross-correlation {$C(k)$} with respect to the ground truth field $\bfz_{\mathrm{truth}}$ using the \texttt{Pylians} \cite{2018ascl.soft11008V} package (see App.~\ref{appendix:summaries} for definitions). We observe an agreement in the power spectrum with the ground truth at $\lesssim 1-2\%$ accuracy for all the scales up to the Nyquist scale $k_{\mathrm{Nyq}} \simeq 0.4 \, h/\mathrm{Mpc}$, and a high ($\gtrsim 50\%$) cross-correlation up to $k \simeq 0.35 \, h/\mathrm{Mpc}$. Furthermore, in order to verify that the distribution of samples correctly covers the true posterior, we perform a Bayesian coverage test for different ranges of $k$-values (see App.~\ref{appendix:coverage}). 

\section{Discussion and conclusions}
\label{discussion}

We have presented and implemented a simple yet precise and effective method for Bayesian field-level reconstruction of cosmological initial conditions. The simplicity of the method lies in the fact that the posterior is modeled to be Gaussian with the covariance being diagonal in Fourier space and the mean determined by a non-linear neural network-based estimator $\mathbf{\hat{\boldmu}}_\bftheta(\bfx_{\mathrm{obs}})$. This turns out to be sufficient to produce accurate results in the chosen range of scales and, after training, the estimator $\mathbf{\hat{\boldmu}}_\bftheta$ and the covariance $(\bm Q_\bftheta^L)^{-1}$ allow us to produce posterior samples in a remarkably fast way. The method does not require the simulator to be differentiable and is not reliant on approximation schemes. 

Our method leaves a lot of promising future directions for investigation. First, we observe that the trained $\bm Q_\bftheta^L$ matrix contains a relatively small amount of scatter and can be approximately described as a function of the wavenumber $k$, as expected from the physical rotational symmetry (see App.~\ref{appendix:map_and_std}). This should make it possible to impose the constraint $\bm Q_\bftheta^L(\bm k) = \bm Q_\bftheta^L(k)$ from the start, significantly reducing the number of trainable parameters and strengthening the interpretability of the approach. Furthermore, since our proposed method is SBI-based, it has the ability to marginalize over cosmological parameters or, alternatively, to infer them alongside the random phases of the ICs field. Especially interesting enhancement in this context can be offered by sequential SBI techniques. 

It is important to realize the present limitations of our approach. While we found the Gaussian diagonal approximation for the posterior to be sufficient in the chosen range of scales $k \lesssim 0.4 \, h/\mathrm{Mpc}$, extending it to smaller scales might require a more complex modeling of the $\bm Q_\bftheta^L$ matrix or to abandon the Gaussian approximation altogether. Moreover, we demonstrate our method in the idealized scenario where the observation consists of the full 3D matter overdensity field. In reality, what we observe are galaxies— 
biased tracers of this field—and additionally, the observations are subject to systematic effects such as redshift space distortions, lightcone effects, survey masks, and noise contamination. While the presented SBI framework is well suited to incorporate all these effects, any realistic application to observational data would require further robustness checks.

\bibliography{references}

\clearpage
\appendix

\section{Appendix: MAP estimation and standard deviation}
\label{appendix:map_and_std}

\begin{figure}
    \centering
    \resizebox{0.8\textwidth}{!}{\includegraphics{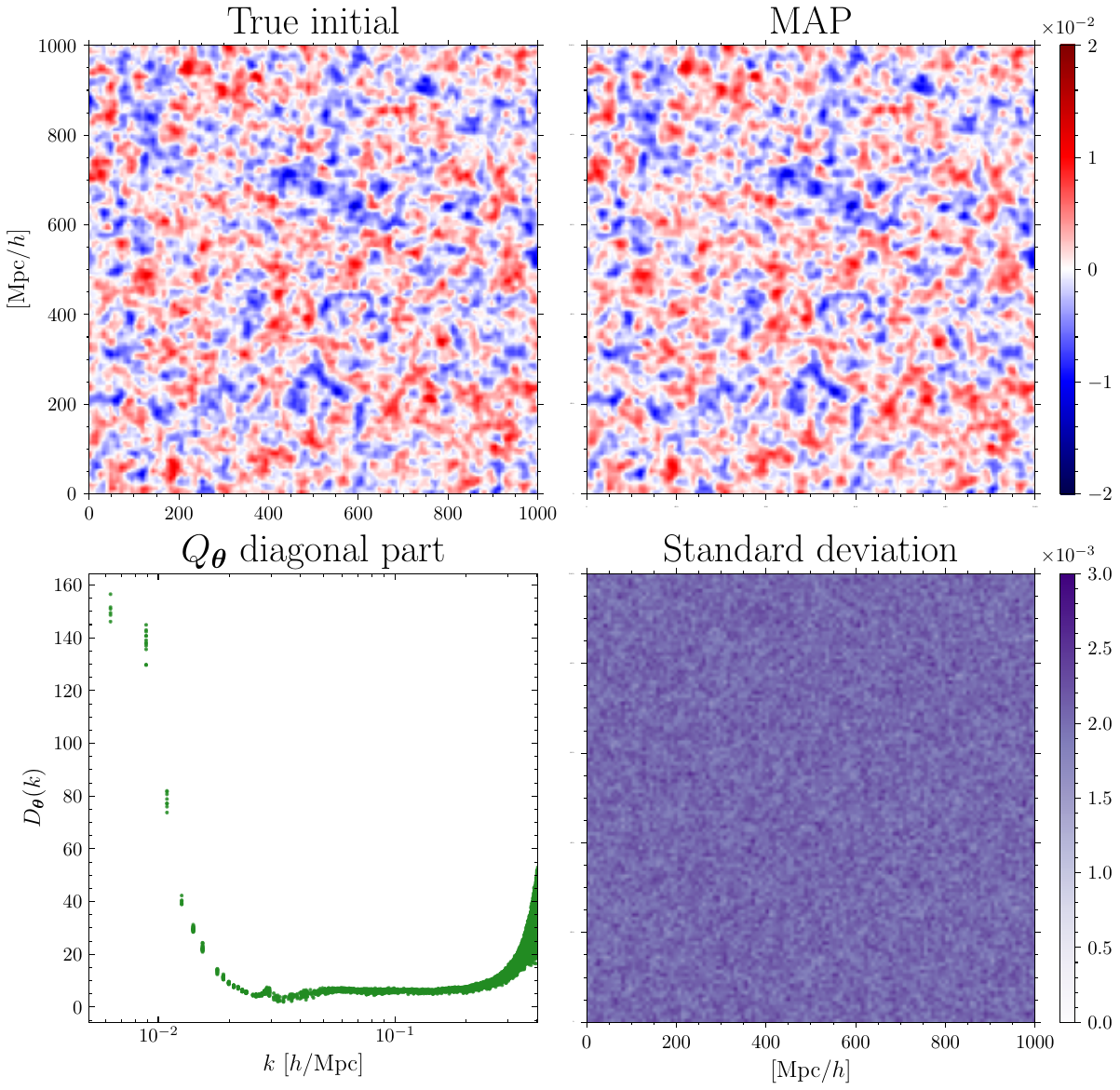}}
    \caption{\textit{Top row}: slices of the initial density field $\bfz_{\rm truth}$ of the target simulation and the corresponding MAP estimate $\mathbf{\hat{\boldmu}}_\bftheta(\bfx_{\mathrm{obs}})$.  \textit{Bottom row left}: diagonal values $\bm D_{\bftheta}$ of the trained posterior precision matrix $\bm Q_\bftheta = \bm Q^P + \bm Q_\bftheta^L$ as a function of the wavenumber $k$. \textit{Bottom row right}: the standard deviation computed from 1000 samples. All the shown slices are averaged over the depth of $100 \ \text{Mpc}/h$ in the third axis direction.}
    \label{fig:map_abd_std}
\end{figure}

In Fig.\ \ref{fig:map_abd_std} we plot the MAP estimate $\mathbf{\hat{\boldmu}}_\bftheta(\bfx_{\mathrm{obs}})$, the standard deviation computed from 1000 samples, and the diagonal values $\bm D_{\bftheta}$ of the trained posterior precision matrix $\bm Q_\bftheta = \bm Q^P + \bm Q_\bftheta^L$ as a function of the wavenumber $k$. The latter quantity allows estimating the precision in the reconstruction at different scales. As expected, we find that $\bm D_{\bftheta}$ is larger for the smallest $k$ values (where evolution is linear and easier to reconstruct), and then it sharply drops to a plateau that extends up to $k_{\mathrm{Nyq}}$.  In addition, we see that $\bm D_{\bftheta}$ shows little variation with the direction of $\bfk$ and can be approximately described as a function of $k$. Hence, knowledge of this function $\bm D_{\bftheta}(k)$ allows to transform any IC point estimator into a fast sampler. Because of this property, the obtained samples for a given $\bfx_{\rm obs}$ have the form of a MAP estimate with a Gaussian random field added on top of it, which causes the standard deviation of samples to be nearly constant (as can be seen in Fig.\ \ref{fig:map_abd_std}). 

In this work, we found that a very simple Fourier-diagonal $\bfx$-independent precision matrix is enough to get consistent posterior samples. However, it might be possible to improve our results by also modeling the $\bm Q_\bftheta$ matrix as $\bfx$-dependent, just as for the $\mathbf{\hat{\boldmu}}_\bftheta$ estimator. For other scenarios, one could also consider alternative structures of the $\bm Q_\bftheta$ matrix. As an example, we experimented with the convolutional matrix $\bm Q_\bftheta^L = \mathcal{C}_{\bftheta}^{\dagger} \bm D_{\bftheta}^L \mathcal{C}_{\bftheta}$, where $\bm D_{\bftheta}^L$ is diagonal and $\mathcal{C}_{\bftheta}$ denotes the convolution operation of a certain kernel size, achieving altogether similar reconstruction results. Instead of convolutions, one can also consider wavelets, scattering transform, etc. In those cases, the prior and the likelihood precision matrices are diagonal in different bases, so sampling is no longer as straightforward and fast, and one needs to resort to techniques such as data augmentation approaches for sampling (see Appendix~B of \cite{List:2023jwo}).

\section{Appendix: Summary statistics}
\label{appendix:summaries}

Given a (dimensionless) 3D matter overdensity field $\delta(\mathbf{x})=\rho(\mathbf{x})/\bar{\rho}-1$ at some fixed redshift, its power spectrum $P(k)$ is defined as follows
\begin{equation}
\langle \tilde{\delta} (\bfk) \tilde{\delta}^{*} (\bfk') \rangle = (2\pi)^3 \, P(k) \, \delta^{(3)}(\bfk - \bfk'),
\label{eq:Pk_def}
\end{equation}
where $\tilde{\delta}(\bfk) = \mathcal{F}\{\delta(\mathbf{x})\}$ denotes the Fourier transform of the field, the brackets $\langle ...\rangle$ indicate the average over many realizations, and the Dirac delta function follows from translational invariance. The power spectrum is the Fourier transform of the matter field two-point correlation function, and thus describes the density fluctuations of the Universe as a function of scale $k$.

Transfer function and cross-correlation between the two fields $\delta_a(\mathbf{x})$ and $\delta_b(\mathbf{x})$ are then defined as:
\begin{equation}
    T_{a b}(k)=\sqrt{\frac{P_a(k)}{P_b(k)}} ; \quad C_{a b}(k)=\frac{P_{a b}(k)}{\sqrt{P_a(k) \times P_b(k)}},
\end{equation}
where the cross-power spectrum $P_{a b}(k)$ is defined analagously to (\ref{eq:Pk_def}) from $\langle \tilde{\delta}_a (\bfk) \tilde{\delta}_b^{*} (\bfk') \rangle$. So, if the two fields are identical, $T_{a b}(k)$ and $C_{a b}(k)$ are equal to $1$ for all $k$, and the deviation of these quantities from $1$ provides a measure of discrepancy in terms of their amplitudes and phases, respectively, between the two fields at different scales $k$.

\section{Appendix: Bayesian coverage test}
\label{appendix:coverage}

We perform a coverage test to evaluate {whether the IC samples generated by our method are statistically consistent with the true posterior distribution, individually for different $k$-bins.} Specifically, we compute the difference between the Fourier 
{transformed} true ICs, $\tilde{\bfz}_\mathrm{truth} := \mathcal{F}\{\bfz_\mathrm{truth}\}$, and the {means of the Fourier transformed}
posterior samples, $\bar{\tilde{\bfz}}_\mathrm{samples} := \mathbb{E}[\tilde{\bfz}_\mathrm{samples}]$, then normalize this difference by the standard deviation of the Fourier transforms of the posterior samples, $\tilde{\sigma}_\mathrm{samples} := \sqrt{\mathbb{E}[(\tilde{\bfz}_\mathrm{samples})^2] - \left(\mathbb{E}[\tilde{\bfz}_\mathrm{samples}]\right)^2}$,
with $\bfz_\mathrm{samples}\sim p(\bfz| \bfx)$ {as modeled in Eq.\ (\ref{posterior})}. 
Mathematically, this can be expressed {via the test statistic}
\begin{equation}
    \label{coverage}
    \Delta(\bm k) := \cfrac{
\tilde{\bfz}_\mathrm{truth} (\bm k) - \bar{\tilde{\bfz}}_\mathrm{samples} (\bm k)}{\tilde{\sigma}_\mathrm{samples} (\bm k)} \;.
\end{equation}

{Since our likelihood precision matrix is highly rotationally symmetric (as expected from the isotropy of the Universe), we focus on the radial dimension in this analysis and consider the consistency of our posterior within different spherical shells in $k$-space.}

\begin{figure}[h!]
    \centering
    \resizebox{0.9\textwidth}{!}{\includegraphics{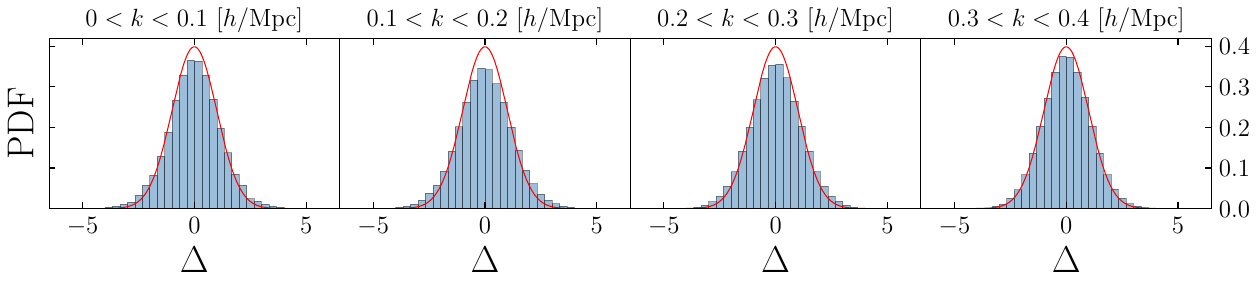}}
    \caption{\textit{Coverage test.} We plot the histograms of the {quantity $\Delta(\bm k)$} defined in (\ref{coverage}) for four different spherical shells in $k$-space. The red lines represent the univariate normal distribution $\mathcal{N}(0, 1)$. The fact that all the histograms closely follow the $\mathcal{N}(0, 1)$ distribution validates our posterior coverage test for different ranges of $k$.
    }
    \label{fig:coverage}
\end{figure}

In Fig.\ \ref{fig:coverage}, we present histograms of these normalized differences $\Delta(\bm k)$ across various $k$-ranges.
The fact that all histograms closely follow the univariate normal distribution with unit variance and zero mean indicates that the posterior samples accurately capture the first two moments of the true posterior distribution.
Note that this is a more challenging test than the one shown in \cite[Fig.\ 5]{legin2023posterior}, where the coverage is 
assessed for all modes at once, {since we demonstrate the statistical consistency of our posterior samples {\it within each individual $k$-bin}.}

\end{document}